\documentclass[%
 reprint,
 amsmath,amssymb,
 aps,
 longbibliography,
 superscriptaddress
]{revtex4-2}

\usepackage{graphicx}
\usepackage{bm}
\usepackage{physics}
\usepackage{hyperref}
\usepackage{xcolor}
\usepackage{booktabs}  

\begin{document}

\title{Measurement-defined control of single-particle interference}

\author{Tai Hyun Yoon}
\email{thyoon@korea.ac.kr}
\affiliation{Department of Physics and Center for Molecular Spectroscopy and Dynamics,
Institute for Basic Science, Korea University, Seoul 02841, Republic of Korea}

\begin{abstract}
Interference is conventionally attributed to path-accumulated phase differences,
with measurement treated as a passive readout.
Here we demonstrate that single-particle interference is governed by the relative
phase between the prepared quantum state and the detector-defined measurement
basis --- a joint quantity that is operationally inaccessible in any conventional
interferometer.
Using coherently seeded entangled nonlinear biphoton sources, we show that
independently scanning the pump phase difference, the seed phase difference,
or the signal interferometric phase each produces identical sinusoidal fringes
($V\approx0.99$) --- a three-scan equivalence impossible in any two-mode
interferometer.
The fringe visibility is continuously controlled through the idler-state overlap,
directly encoding quantum distinguishability without idler detection.
The same measurement-defined interference law persists from the single-photon to
the high-flux regime.
The bright--dark collective-state structure demonstrated here unifies coherent
population trapping and electromagnetically induced transparency in atomic
$\Lambda$-systems, discrete photonic interference, and single-slit diffraction
within a common framework differing only in dark-subspace dimensionality,
establishing measurement-defined photonic modes as a fundamental resource for
quantum photonics.
\end{abstract}

\maketitle


Interference is one of the central phenomena of wave physics~\cite{Young1802}
and quantum physics~\cite{Colella1975}, and is conventionally understood as arising
from the superposition of propagation paths~\cite{Dirac1958,Bohr1928,Feynman1965}.
This paradigm underlies a wide range of experiments, including single-particle
interference of electrons, atoms, neutrons, and photons~\cite{Tonomura1989,Riehle1991}.
In quantum theory, however, the detection process is not a passive readout but an
active interaction~\cite{Heisenberg1927,Zurek2003}: photodetection is fundamentally governed by the coupling between
the electromagnetic field and the detector, as formalized by Glauber~\cite{Glauber1963}.
Within this framework, detection corresponds to projection onto specific field modes
defined by the detector--field coupling operator $\hat{O}$, whose eigenmodes form
orthogonal \emph{bright} and \emph{dark} collective states satisfying
$\hat{O}|D\rangle=0$~\cite{VillasBoas2025}.
A quantum system in the dark state carries energy or photons yet remains entirely
invisible to the detector, while only the bright-state component produces a measurable
signal.
This measurement-centered perspective suggests that interference minima reflect
population of dark states, physically occupied but undetectable modes, rather
than the absence of the quantum.

The bright--dark collective-state structure is not unique to any single physical
system but constitutes a universal principle that has been identified across atomic and
photonic platforms.
In three-level atomic $\Lambda$-systems, coherent population trapping (CPT) and
electromagnetically induced transparency (EIT) are its oldest and best-established
manifestations~\cite{Arimondo1996,Aspect1988}: the coupling operator
$\hat{V}=\frac{1}{\Omega}(\Omega_1|e\rangle\langle g_1|+\Omega_2|e\rangle\langle g_2|), \Omega = \sqrt{\Omega_1^2+\Omega_2^2},$ defines a dark
ground-state superposition $|D\rangle_\Lambda$ that satisfies
$\hat{V}|D\rangle_\Lambda=0$, directly forbidding photon absorption through
destructive interference of the two excitation pathways, leaving the excited state
unpopulated and suppressing fluorescence as a further consequence.
In the photonic domain, Villas-Boas~\emph{et al.}~\cite{VillasBoas2025} recently
extended Glauber's photodetection theory to recast double-slit and multi-slit
interference in the same terms: the positive-frequency operator
$\hat{E}^{(+)}=\frac{1}{\sqrt{2}}(\hat{a}_1+e^{i\phi}\hat{a}_2)$ at a given
detector position defines one bright and one dark photonic mode, and the double-slit
fringe depending on local phase $\phi = 2n\pi, n\in\mathbb{Z},$ is identified with the bright-mode occupation probability, while photons at
interference minima reside in the dark mode rather than being absent with $\phi = 2(n+1/2)\pi$.
From the complementary source perspective, Qian and Agarwal~\cite{Qian2020} showed
that the purity of the quanton source itself bounds the totality of wave-particle
duality through the relation $P^2+V^2=\mu_s^2$, where $P$ is the path predictability,
$V$ is the fringe visibility, and $\mu_s$ is the source purity, establishing that
interference visibility is limited not only by the measurement geometry but by the
degree of entanglement inherent in the source that generates the quanton.
While the Villas-Boas~\emph{et al.}\ framework addresses how the detector coupling
defines which modes are visible, the Qian--Agarwal framework addresses how the source
properties limit the quanton's wave character: both perspectives are unified in
the present ENBS system, where the idler-state fidelity $F=|\langle I_1|I_2\rangle|$
directly plays the role of $\mu_s$, and source and detector degrees of freedom are
independently accessible~\cite{Yoon2021}.
Cheng~\emph{et al.}~\cite{Cheng2025} further extended this discrete-mode framework
to the continuous-mode regime of single-slit Fraunhofer diffraction, where the
detector-oriented Fourier basis requires an infinite-dimensional dark subspace, the
classical sinc profile $(\sin\beta/\beta)^2$ emerges as the bright-mode population,
and $N$-photon Fock states are predicted to exhibit a spatially uniform
second-order correlation $G^{(2)}=1-1/N$ as a distinctive nonclassical signature.
These four systems (atomic CPT/EIT, discrete photonic interference, the present
experiment, and continuous-mode diffraction) are all governed by the same
collective-state structure, as summarized in Table~\ref{tab:comparison}, differing
only in the dimensionality of the dark subspace and the physical origin of the
coupling operator; the Qian--Agarwal source-purity framework~\cite{Qian2020}
provides a complementary description of the same physics from the emitter side.

Despite this theoretical convergence, direct experimental access to the
bright--dark mode picture has remained limited.
In the atomic $\Lambda$-scheme, the same laser fields simultaneously define the
coupling Hamiltonian and drive the atomic state, so the bright/dark basis and the
prepared state cannot be varied independently.
In conventional photonic interferometers, both the prepared state (determined by the
propagation geometry from the source) and the measurement basis (fixed by the
detector position) are set by the same physical geometry, providing no intrinsic
handle on which-path distinguishability.
Which-path information, when introduced at all, is typically imposed externally
through path marking or entanglement with auxiliary systems, as in quantum eraser
experiments~\cite{Scully1982,Kim2000,Jacques2007,Tang2012,Ma2013}.
As a result, although the bright--dark mode picture is well established theoretically,
it has not been realized in a system where the prepared quantum state and the
detector-defined measurement basis can be independently and continuously controlled
within a single device.

Here we realize such a system using a hybrid time--frequency interferometer based on
coherently seeded entangled nonlinear biphoton sources
(ENBSs)~\cite{Lee2018,Yoon2021}, constituting the first direct experimental
demonstration of the Villas-Boas~\emph{et al.} framework~\cite{VillasBoas2025}.
The experimental signature specific to this demonstration is the three-scan
equivalence: scanning $\Delta\phi_s$ (signal interferometric phase, which rotates
the measurement basis), $\Delta\phi_p\equiv\phi_{p2}^{(\mathrm{in})}-\phi_{p1}^{(\mathrm{in})}$
(pump input phase difference), or
$\Delta\phi_{sd}\equiv\phi_{sd2}^{(\mathrm{in})}-\phi_{sd1}^{(\mathrm{in})}$
(seed input phase difference) independently all produce identical sinusoidal
fringes, demonstrating that interference is governed solely by the relative phase
$(\phi+\phi_s)$ between the prepared state and the detector-defined basis, a
result that is impossible in any conventional interferometer where state preparation
and measurement basis are set by the same propagation geometry.
In this platform, a single-photon signal mode is intrinsically correlated with
auxiliary idler states that encode which-path information, enabling simultaneous and
continuous control of interference and distinguishability through internal quantum
correlations, with the state preparation phases $\phi=\Delta\phi_{sd}-\Delta\phi_p$
and measurement phase $\phi_s=k_s\Delta x_s$ independently controlled through
physically separate degrees of freedom.
The visibility of the signal single-photon interference fringe is quantitatively determined by the
idler-state overlap $F=|\langle I_1|I_2\rangle|$, providing a direct operational
link between coherence and distinguishability, while the same detector-defined modal
structure persists from the single-photon regime to high-flux operation.

\begin{figure*}[t]
\centering
\includegraphics[width=0.85\linewidth]{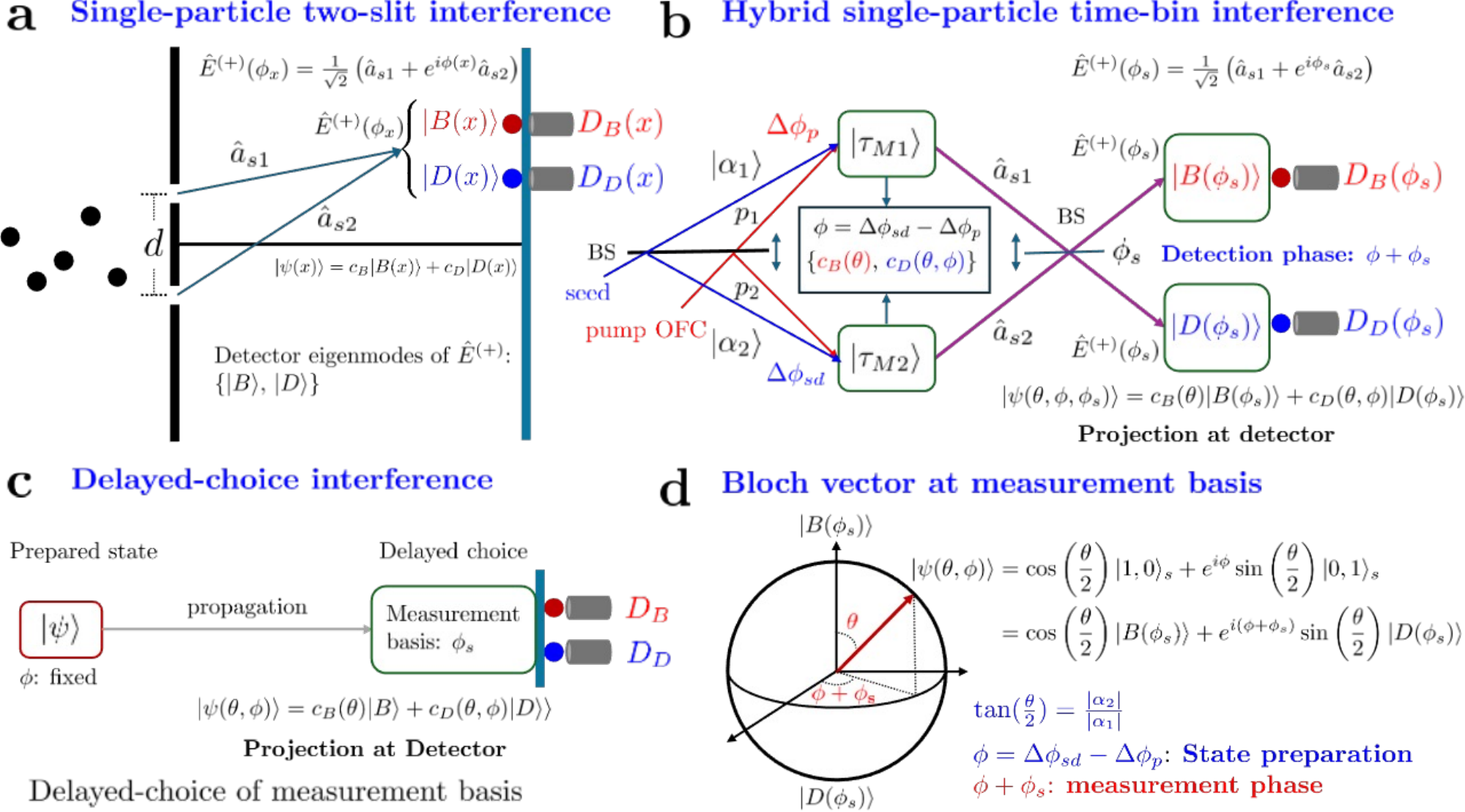}
\caption{\textbf{Detector-defined photonic states
governing interference across space and time.}
\textbf{a}, Spatial two-alternative interference
viewed as projection onto detector-defined collective
photonic states.
The normalized detector--field coupling operator
$\hat{E}^{(+)}(\phi_x)=\frac{1}{\sqrt{2}}(\hat{a}_{s1}+e^{i\phi(x)}\hat{a}_{s2})$
defines orthogonal eigenmodes $\{|B(x)\rangle,|D(x)\rangle\}$
at detection coordinate $x$, onto which the single-quantum state
$|\psi(x)\rangle=c_B|B(x)\rangle+c_D|D(x)\rangle$
is projected at detector $D_j$.
\textbf{b}, Time--frequency (TF) realization using
coherently seeded entangled nonlinear biphoton sources (ENBSs).
At each time bin $\tau_M$, the pump pulse is split into two
independent ENBS modules, generating signal modes $|\tau_{M1}\rangle$
and $|\tau_{M2}\rangle$ from the same pump time bin.
The ENBS input phase differences
$\Delta\phi_{sd}\equiv\phi_{sd2}^{(\mathrm{in})}-\phi_{sd1}^{(\mathrm{in})}$
and $\Delta\phi_p\equiv\phi_{p2}^{(\mathrm{in})}-\phi_{p1}^{(\mathrm{in})}$ determine the
prepared state $|\psi(\theta,\phi)\rangle$ with coefficients
$\{c_B(\theta),c_D(\theta,\phi)\}$, azimuthal phase
$\phi=\Delta\phi_{sd}-\Delta\phi_p$, and polar angle
$\tan(\theta/2)=|\alpha_2|/|\alpha_1|$.
The signal phase $\phi_s$ at the output beam splitter rotates the
measurement basis, defined by the normalized operator
$\hat{E}^{(+)}(\phi_s)=\frac{1}{\sqrt{2}}(\hat{a}_{s1}+e^{i\phi_s}\hat{a}_{s2})$,
projecting onto the detector eigenmodes $|B(\phi_s)\rangle$ and
$|D(\phi_s)\rangle$ with detection phase $\phi+\phi_s$.
\textbf{c}, Delayed-choice control.
A late choice of $\phi_s$ rotates the measurement basis without
modifying the prepared state $|\psi(\theta,\phi)\rangle$,
implementing Wheeler's delayed-choice principle in a mode-resolved
TF platform.
\textbf{d}, Bloch vector at measurement basis.
The prepared signal state is expressed in both the computational basis
$|\psi(\theta,\phi)\rangle=\cos(\theta/2)|1,0\rangle_s
+e^{i\phi}\sin(\theta/2)|0,1\rangle_s$
and the $\phi_s$-dependent measurement basis
$=\cos(\theta/2)|B(\phi_s)\rangle+e^{i(\phi+\phi_s)}\sin(\theta/2)|D(\phi_s)\rangle$,
where the north and south poles $|B(\phi_s)\rangle$ and $|D(\phi_s)\rangle$
are the detector-defined eigenmodes for the given $\phi_s$.
The polar angle $\theta=2\arctan(|\alpha_2|/|\alpha_1|)$ is set by the
seed amplitude ratio and controls the fringe visibility; the azimuthal
angle $\phi=\Delta\phi_{sd}-\Delta\phi_p$ is the state preparation phase (blue),
controlled by the StPDC phase-matching condition independently of $\phi_s$.
The measurement phase $\phi+\phi_s$ (red) is the azimuthal angle of the
Bloch vector in the measurement-basis frame:
rotating $\phi_s$ rotates the poles of the Bloch sphere without changing
the Bloch vector $(\theta,\phi)$, so that interference is governed by
the detection phase $\phi+\phi_s$.}
\label{fig:fig1}
\end{figure*}

Our implementation employs a time--frequency interferometer driven by stabilized
optical frequency-comb pumps, which offer phase-coherent modes across a broad
spectral bandwidth (Methods)~\cite{Udem2002,Cundiff2003} and support high-dimensional photonic
states for time--frequency quantum information
processing~\cite{Kues2019,Reimer2014,Fujimoto2022,Yamazaki2022}.
As illustrated in Fig.~\ref{fig:fig1}, detection at the output beam splitter
corresponds to projection onto bright and dark eigenmodes of the detector--field
coupling [Fig.~\ref{fig:fig1}\textbf{a}], the prepared state is parameterized by the Bloch
sphere $|\psi(\theta,\phi)\rangle=\cos(\theta/2)|B(\phi_s)\rangle+e^{i(\phi+\phi_s)}\sin(\theta/2)|D(\phi_s)\rangle$
[Fig.~\ref{fig:fig1}\textbf{d}], and the measurement basis is rotated independently by
$\phi_s$ [Fig.~\ref{fig:fig1}\textbf{b,c}].

\section*{Theory}

Two coherently seeded entangled nonlinear biphoton sources (ENBSs) are driven by
a stabilized optical frequency comb, each converting spontaneous parametric
downconversion (SPDC) into stimulated parametric downconversion (StPDC) upon
injection of a seed field at the idler frequency $\omega_i$~\cite{Lee2018,Yoon2021}.
The StPDC phase-matching condition for ENBS$_j$ is
\begin{equation}
    \Phi_j = \phi_{pj}^{(\mathrm{in})} - \bigl(\phi_{sj}^{(\mathrm{out})} + \phi_{sdj}^{(\mathrm{in})}\bigr) = 0,
    \label{eq:pmatch}
\end{equation}
which directly fixes the signal output phase
$\phi_{sj}^{(\mathrm{out})}=\phi_{pj}^{(\mathrm{in})}-\phi_{sdj}^{(\mathrm{in})}$
in terms of the controllable pump and seed phases, in contrast to unseeded SPDC
where the idler phase is vacuum-seeded and uncontrolled.
Expanding the time-evolved state to first order in the weak-gain parameter
$r_j\equiv|\kappa_j|t_j/\hbar\ll 1$ (where $t_j=n_pL/c$ is the pump transit
time) and tracing over the idler modes (see Supplementary Note~1 for the full
derivation including the SPACS idler structure and QPM condition) yields the
reduced signal density matrix in the basis $\{|1,0\rangle_s,|0,1\rangle_s\}$.
The complete PPLN type-0 coupling constant for the 530\,nm\,$\to$\,807\,nm\,$+$\,1542\,nm
process is
\begin{equation}
\kappa_j = \frac{\hbar\,d_\mathrm{eff}^{(0)}}{n_p}
  \sqrt{\frac{P_{p,j}\,\omega_s\,\omega_i}
             {\epsilon_0\,n_p\,n_s\,n_i\,c\,A_\mathrm{eff}}}\;
  \mathrm{sinc}\!\left(\frac{\Delta k_j\,L}{2}\right)
  e^{i\phi_{pj}^{(\mathrm{in})}},
\label{eq:kappa}
\end{equation}
where $d_\mathrm{eff}^{(0)}=(2/\pi)d_{33}$ for first-order type-0 PPLN poling,
$n_p$, $n_s$, $n_i$ are the extraordinary refractive indices at 530, 807, and
1542\,nm respectively, $P_{p,j}$ is the pump power, $A_\mathrm{eff}$ is the
effective mode area, and $\Delta k_j$ is the quasi-phase-mismatch defined in
Supplementary Note~1 [Eq.~(S2)].
The density matrix
\begin{equation}
\rho_s =
\begin{pmatrix}
\rho_{11} & \sqrt{\rho_{11}\rho_{22}}\,F\,e^{-i\phi} \\
\sqrt{\rho_{11}\rho_{22}}\,F\,e^{i\phi} & \rho_{22}
\end{pmatrix},
\label{eq:rho}
\end{equation}
where $\rho_{jj} = r_j^2|\alpha_j|^2$ are the diagonal populations
(with $r_j$ the squeezing parameter of Supplementary Note~1 [Eq.~(S3)]),
the off-diagonal phase is
\begin{equation}
\phi = \Delta\phi_{sd} - \Delta\phi_p,
\label{eq:phi_def}
\end{equation}
and $F$ is the idler-state overlap defined below.

\begin{figure*}[t]
\centering
\includegraphics[width=0.85\linewidth]{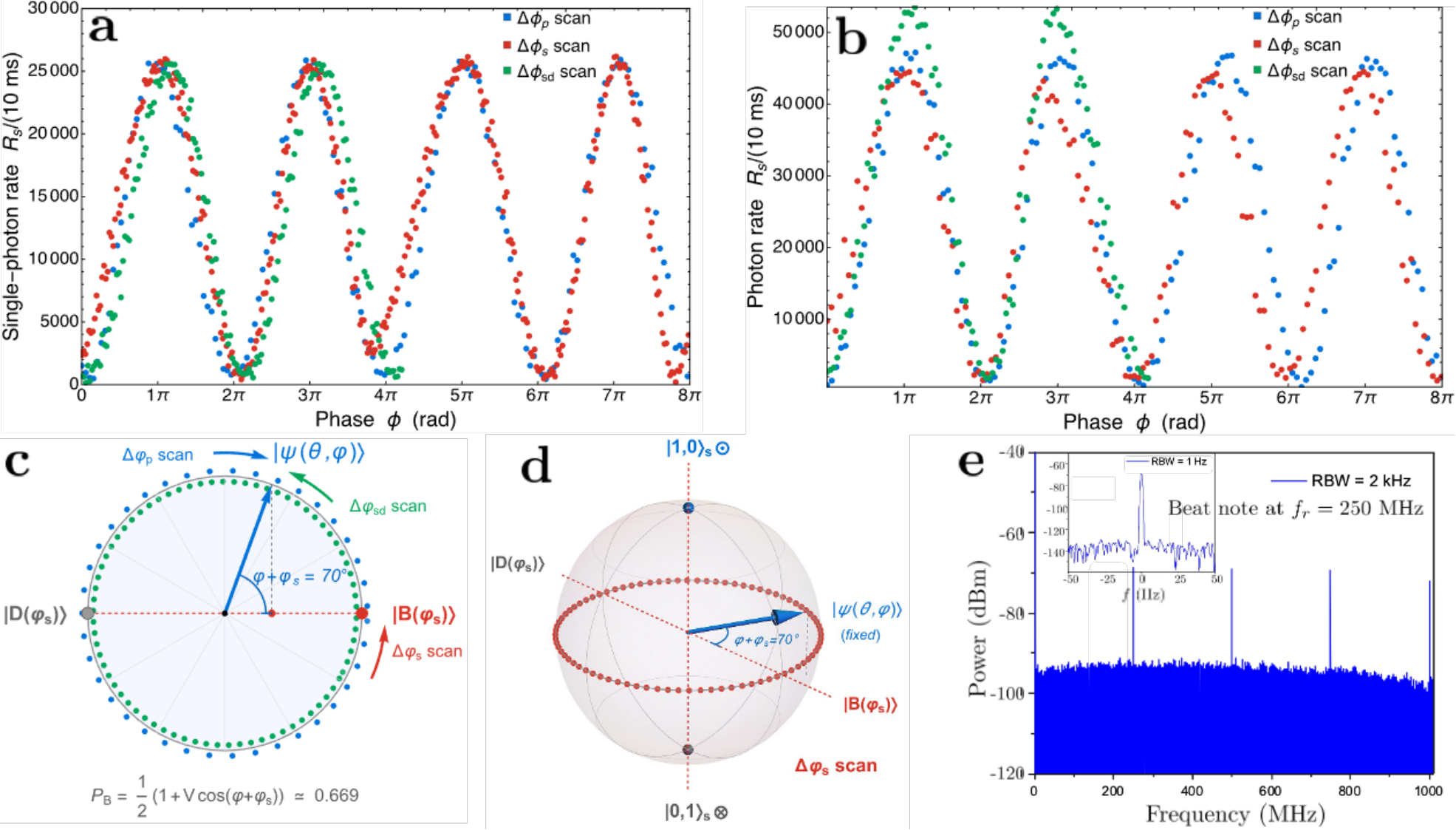}
\caption{\textbf{Experimental demonstration of measurement-defined interference
across quantum and classical regimes.}
\textbf{a}, Single-photon interference fringes, each plotted against its own
independently scanned phase parameter $\Delta\phi_j$ (rad): pump phase difference
$\Delta\phi_p$ (blue), signal interferometric phase $\Delta\phi_s$ (red), and seed
phase difference $\Delta\phi_{sd}$ (green).
The $y$-axis shows the single-photon count rate $R_s$ per 10\,ms integration
frame detected by the EMCCD.
All three datasets exhibit sinusoidal modulation with the same period ($2\pi$)
and visibility, consistent with $P_B=\frac{1}{2}[1+V\cos(\phi+\phi_s)]$
[Eq.~(\ref{eq:brightP})], confirming that interference is governed by the
relative phase between state preparation and measurement regardless of which
phase parameter is scanned.
\textbf{b}, High-flux interference fringes under identical phase control,
showing the same sinusoidal phase law at a photon rate $\sim$2$\times$ larger
than the single-photon regime, demonstrating that the detector-defined modal
structure and measurement-defined phase law persist beyond the single-photon
regime.
\textbf{c}, Equatorial Bloch-sphere representation of the single-photon data
from panel~\textbf{a} for the $\Delta\phi_p$ (blue, outer ring at $r=1.04$)
and $\Delta\phi_{sd}$ (green, inner ring at $r=0.96$) scans.
Both datasets uniformly cover the complete unit circle, confirming that the
Bloch vector traces the full equator ($\theta=\pi/2$, $V\approx0.99$) and
that only the detection phase $\phi+\phi_s$ governs $P_B$.
The snapshot angle $\phi+\phi_s=70^\circ$ and the resulting $P_B\approx0.669$
are indicated.
\textbf{d}, Three-dimensional Bloch sphere for the $\Delta\phi_s$ scan from
panel~\textbf{a}, showing the Bloch vector fixed in the equatorial plane while
the measurement basis $\{|B(\phi_s)\rangle,|D(\phi_s)\rangle\}$ (red dashed
axes) rotates.
Red dots are the experimental data mapped onto the unit equatorial circle
($r=1$, $z=0$); north and south poles correspond to the computational basis
$|1,0\rangle_s$ (out of page, $\odot$) and $|0,1\rangle_s$ (into page,
$\otimes$), respectively.
\textbf{e}, Radio-frequency beat-note spectrum of the high-flux output field.
Main panel: broad scan (0--1000\,MHz, RBW\,=\,2\,kHz) showing comb lines at
harmonics of the repetition rate $f_r=250$\,MHz.
Inset: zoomed view centered on the $f_r$ comb line (RBW\,=\,1\,Hz,
frequency axis in Hz), confirming an ultra-narrow linewidth and that the
output forms a phase-coherent pulse train occupying the detector-defined
photonic modes.
Data in panel \textbf{a} replotted from Ref.~\cite{Lee2018} and reinterpreted
within the bright--dark modal framework; data in panels \textbf{b}--\textbf{e}
recorded in the same experimental campaign as Ref.~\cite{Lee2018} but not
previously reported.}
\label{fig:fig2}
\end{figure*}

The coherence of the signal state is governed by the overlap of the idler states,
\begin{equation}
F \equiv |\langle I_1|I_2\rangle|
= \frac{|\alpha_1||\alpha_2|}
{\sqrt{(1+|\alpha_1|^2)(1+|\alpha_2|^2)}},
\label{eq:overlap}
\end{equation}
which provides a continuous measure of quantum distinguishability.
For $|\alpha_j|\ll 1$, the idler states are nearly orthogonal and the signal paths are
distinguishable, while for $|\alpha_j|\gg 1$, the states become indistinguishable and
classical interference is recovered.
In the language of Qian and Agarwal~\cite{Qian2020}, $F$ plays the role of the
source purity $\mu_s$ that bounds the totality of wave-particle duality through
$P^2+V^2=\mu_s^2$: the idler-state overlap directly controls how much wave-like
behavior ($V$) or particle-like behavior ($P$) the signal quanton can exhibit,
linking the source-perspective and detector-perspective descriptions of
complementarity within a single experimentally accessible parameter.
In a separate experimental run in which $|\alpha|^2$ was varied by inserting a
variable neutral density (ND) filter on one seed path, the maximum achievable
visibility was $V \approx 0.89$ rather than unity, because inserting the ND filter
slightly degraded the spatial overlap of the two signal beams at the output beam
splitter~\cite{Lee2018}.
The theoretical prediction in that run is accordingly $V=0.9\times|\alpha|^2/(1+|\alpha|^2)$,
and at $|\alpha|^2\approx 100$ ($|\alpha|\approx 10$) this gives $V\approx 0.89$,
confirming near-classical operation at the upper end of the quantum-to-classical
transition; the full variation of $V$ with $|\alpha|^2$ is demonstrated in
Fig.~3(b) of Ref.~\cite{Lee2018}.
By contrast, in the three phase-scan runs of Fig.~\ref{fig:fig2}\textbf{a}, where no ND filter was inserted and the spatial overlap was optimized, the measured
visibility reaches $V\approx 0.99$, consistent with the ideal prediction
$V=|\alpha|^2/(1+|\alpha|^2)$ at $|\alpha|^2\approx 100$.

Detection is described by the normalized positive-frequency operator
\begin{equation}
\hat{E}_s^{(+)} = \frac{1}{\sqrt{2}}\left(a_{s1} + e^{i\phi_s}a_{s2}\right),
\end{equation}
which satisfies $[\hat{E}_s^{(+)},\hat{E}_s^{(-)}]=1$ and defines the detector
eigenmodes.
The bright state is
\begin{equation}
|B(\phi_s)\rangle = \frac{1}{\sqrt{2}}\left(|1,0\rangle_s + e^{-i\phi_s}|0,1\rangle_s\right).
\end{equation}
The orthogonal dark state is
\begin{equation}
    |D(\phi_s)\rangle = \frac{1}{\sqrt{2}}
    \left(|1,0\rangle_s - e^{-i\phi_s}|0,1\rangle_s\right),
    \label{eq:dark}
\end{equation}
which satisfies $\hat{E}^{(+)}_s|D(\phi_s)\rangle=0$ and is therefore invisible to
the detector.

The prepared signal state can be represented in the detector-defined basis
$\{|B(\phi_s)\rangle,|D(\phi_s)\rangle\}$ as a general Bloch vector
\begin{equation}
    |\psi(\theta,\phi)\rangle
    = \cos\frac{\theta}{2}\,|B(\phi_s)\rangle
    + e^{i(\phi+\phi_s)}\sin\frac{\theta}{2}\,|D(\phi_s)\rangle,
    \label{eq:bloch}
\end{equation}
where the polar angle $\theta$ is determined by the relative amplitudes of the two
signal modes.
In the StPDC process, the signal generation rate of each ENBS$_j$ is stimulated
by the seed field, so the effective coupling amplitude $\eta_j = r_j|\alpha_j|$
(Supplementary Note~1 [Eq.~(S4)]) gives $\rho_{jj} = r_j^2|\alpha_j|^2$.
For equal bare nonlinear coupling ($r_1 = r_2 \equiv r$), the polar angle is therefore
\begin{equation}
    \tan\frac{\theta}{2}
    = \sqrt{\frac{\rho_{22}}{\rho_{11}}}
    = \frac{|\alpha_2|}{|\alpha_1|},
    \label{eq:polar}
\end{equation}
showing that $\theta$ is directly and continuously controlled by the ratio of the two
seed amplitudes $|\alpha_j|$, and hence by the coherent seeding powers.
The azimuthal angle
\begin{equation}
    \phi = \Delta\phi_{sd} - \Delta\phi_p,
    \label{eq:azimuth}
\end{equation}
is the prepared-state phase of Eq.~(\ref{eq:phi_def}),
determined by the StPDC phase-matching
condition Eq.~(\ref{eq:pmatch}) and independently controlled through the pump and seed
path-length differences [Fig.~\ref{fig:fig1}\textbf{d}], where
$\Delta\phi_p=\phi_{p2}^{(\mathrm{in})}-\phi_{p1}^{(\mathrm{in})}$ and $\Delta\phi_{sd}=\phi_{sd2}^{(\mathrm{in})}-\phi_{sd1}^{(\mathrm{in})}$
are the pump and seed phase differences between the two ENBSs.
The two angles $(\theta,\phi)$ parameterize the prepared state on the Bloch sphere
entirely independently of the measurement basis $\phi_s=k_s\Delta x_s$, which rotates
the detector eigenmodes $|B(\phi_s)\rangle$ and $|D(\phi_s)\rangle$ at the output beam
splitter without affecting either $\theta$ or $\phi$.

\begin{table*}[t]
\caption{\textbf{Unified bright--dark collective-state structure across four quantum
systems.}
The common mathematical skeleton (a coupling operator $\hat{O}$, a bright eigenstate satisfying $\hat{O}^\dagger\hat{O}|B\rangle\neq 0$, and a dark eigenstate satisfying $\hat{O}|D\rangle=0$) appears across atomic and photonic systems
spanning discrete and continuous mode spaces.
The degree of independent control over state preparation and measurement
distinguishes the four cases.
$\Omega\equiv\sqrt{\Omega_1^2+\Omega_2^2}$; see text for all other symbols.}
\label{tab:comparison}
\renewcommand{\arraystretch}{1.35}
\footnotesize
\begin{ruledtabular}
\begin{tabular}{p{1.7cm}p{2.9cm}p{2.9cm}p{2.9cm}p{2.9cm}}
\textbf{Element}
  & \textbf{Atomic $\Lambda$-system\newline (CPT/EIT)}
  & \textbf{Discrete photonic\newline interference \cite{VillasBoas2025}}
  & \textbf{TF-ENBS\newline (this work)}
  & \textbf{Single-slit\newline diffraction \cite{Cheng2025}} \\[2pt]
\hline\\[-6pt]
Alternatives
  & Ground states\newline $|g_1\rangle$, $|g_2\rangle$
  & Spatial paths\newline $|s_1\rangle$, $|s_2\rangle$
  & Temporal modes\newline $|\tau_1\rangle$, $|\tau_2\rangle$
  & Slit modes $|1_x\rangle$,\newline $x\in[0,b]$ \\[2pt]
Detector
  & Excited state $|e\rangle$
  & Screen position $x$
  & Output BS\,+\,EMCCD
  & Sensor atom at $\theta$ \\[2pt]
Coupling $\hat{O}$
  & $\hat{V}=\frac{1}{\Omega}(\Omega_1|e\rangle\langle g_1|$\newline$\;\;+\Omega_2|e\rangle\langle g_2|)$
  & $\hat{E}^{(+)}\!=\!\frac{1}{\sqrt{2}}(\hat{a}_1$\newline$\quad+e^{i\phi(x)}\hat{a}_2)$
  & $\hat{E}^{(+)}\!=\!\frac{1}{\sqrt{2}}(\hat{a}_1$\newline$\quad+e^{i\phi_s}\hat{a}_2)$
  & $\hat{E}^{(+)}(\theta)=$\newline$\;\tfrac{1}{\sqrt{b}}\!\int\!\hat{a}(x)e^{i\phi(x,\theta)}dx$ \\[2pt]
Dark condition
  & $\hat{V}|D\rangle_\Lambda=0$
  & $\hat{E}^{(+)}|D(x)\rangle=0$
  & $\hat{E}^{(+)}|D(\phi_s)\rangle=0$
  & $\hat{E}^{(+)}|D_{n,\theta}\rangle=0$,\newline $n\neq 0$ \\[2pt]
Bright $|B\rangle$
  & $(\Omega_1|g_1\rangle$\newline$\;+\Omega_2|g_2\rangle)/\Omega$
  & $(|s_1\rangle$\newline$\;+e^{-i\phi}|s_2\rangle)/\sqrt{2}$
  & $(|\tau_1\rangle$\newline$\;+e^{-i\phi_s}|\tau_2\rangle)/\sqrt{2}$
  & $|\psi_0(\theta)\rangle$ \\[2pt]
Dark $|D\rangle$
  & $(\Omega_2|g_1\rangle$\newline$\;-\Omega_1|g_2\rangle)/\Omega$
  & $(|s_1\rangle$\newline$\;-e^{-i\phi}|s_2\rangle)/\sqrt{2}$
  & $(|\tau_1\rangle$\newline$\;-e^{-i\phi_s}|\tau_2\rangle)/\sqrt{2}$
  & $\{|\psi_n\rangle\}_{n\neq 0}$\newline ($\infty$-dim.) \\[2pt]
Hilbert dim.
  & Qubit (2D)
  & $N$D ($N$ slits)
  & Qubit (2D)
  & Infinite \\[2pt]
Bloch sphere
  & Yes:\newline $\theta\!:\Omega_2/\Omega_1$\newline $\phi\!:\phi_2\!-\!\phi_1$
  & Yes (N=2):\newline geometry fixed
  & Yes:\newline $\theta\!:|\alpha_2|/|\alpha_1|$\newline $\phi\!:\Delta\phi_{sd}\!-\!\Delta\phi_p$
  & No \\[2pt]
Basis rotation
  & Raman detuning\newline $\Delta=\delta_1-\delta_2$
  & Screen position $x$
  & Signal phase\newline $\phi_s=k_s\Delta x_s$
  & Angle $\theta$ \\[2pt]
Indep.\ $(\theta,\phi)$\newline vs.\ basis
  & No\newline (same laser fields)
  & No\newline (propagation geom.)
  & \textbf{Yes}\newline (separate controls)
  & No\newline (geom.\ fixed) \\[2pt]
Observable
  & Absorption \&\newline fluorescence\newline suppressed (EIT/CPT)
  & $P(x)\propto$\newline$(1+\cos\phi(x))/2$
  & $P_B\!=\!\tfrac{1}{2}[1$\newline$\quad+V\cos(\phi\!+\!\phi_s)]$
  & $P(\theta)\propto$\newline$(\sin\beta/\beta)^2$ \\[2pt]
Exp.\ status
  & Demonstrated\newline \cite{Arimondo1996,Aspect1988}
  & Proposed\newline \cite{VillasBoas2025}
  & This work\newline \cite{Lee2018,Yoon2021}
  & Proposed\newline \cite{Cheng2025} \\
\end{tabular}
\end{ruledtabular}
\end{table*}

Interference therefore arises from the projection of $|\psi(\theta,\phi)\rangle$ onto
$|B(\phi_s)\rangle$, governed by the detection phase $(\phi+\phi_s)$ between the
prepared state and the measurement basis: this quantity is inaccessible in
conventional interferometers where state preparation and measurement cannot be
independently controlled.
The detection probability is therefore
\begin{equation}
    P_B = \langle B(\phi_s)|\rho_s|B(\phi_s)\rangle
    = \frac{1}{2}\left[1 + V\cos(\phi + \phi_s)\right],
    \label{eq:brightP}
\end{equation}
where $\phi=\Delta\phi_{sd}-\Delta\phi_p$ [Eq.~(\ref{eq:azimuth})] is the prepared-state
phase set by the StPDC phase-matching condition, and
$V=2\sqrt{\rho_{11}\rho_{22}}\cdot F$ is the fringe visibility governed by the
idler-state overlap $F$ [Eq.~(\ref{eq:overlap})].
The signal interferometric phase $\phi_s$, by contrast, is introduced solely at the
output beam splitter through the path-length difference $\Delta x_s$ at the signal
wavelength, $\phi_s=k_s\Delta x_s$, and acts exclusively as a rotation of the
detector measurement basis, with no effect on the prepared state $\phi$.

In this experiment, the idler-state overlap $F=|\langle I_1|I_2\rangle|$ [Eq.~(\ref{eq:overlap})]
and the diagonal populations $\rho_{11}=r_1^2|\alpha_1|^2$,
$\rho_{22}=r_2^2|\alpha_2|^2$ (with $r_j$ the SPDC squeezing parameter of
Supplementary Note~1) are determined entirely by the input seed amplitudes
$|\alpha_j|$ and pump powers $P_{p,j}$, without any detection of the idler output.
Consequently, the interference visibility $V=2\sqrt{\rho_{11}\rho_{22}}\cdot F$ is
continuously and intrinsically tunable by adjusting the seed powers alone~\cite{Yoon2021},
with no ancillary idler measurement required.

The same formalism can be extended to include interferometric measurements on the
idler modes as a separate experimental capability.
By introducing a beam splitter in the idler paths, the overlap $\langle I_1|I_2\rangle$
can be accessed through joint signal--idler correlation measurements, enabling quantum
eraser--type experiments in which which-path information is erased or recovered by
conditioning on idler detection.
In practice, since each SPACS idler state $|I_j\rangle$ is produced on top of the
residual CW seed $|\alpha_j\rangle$ with $|\alpha_j|^2\gg 1$ seed photons per pulse,
isolating the single stimulated photon from the bright seed background requires
signal--idler coincidence postselection, which automatically suppresses uncorrelated
seed photons and retains only events where an idler detection is correlated with signal
photon generation; equivalently, time-gated single-photon detection synchronized to
the pump pulse window could suppress the CW background by the ratio of the gate
width to the comb period.
This would provide direct experimental access to the bipartite structure of the state
and link interference to measurable quantum correlations, extending the present
platform to the full quantum eraser regime.

\section*{Unified collective-state structure across four quantum systems}

The bright--dark decomposition is an instance of a general
collective-state principle that appears in physically distinct quantum systems.
Table~\ref{tab:comparison} summarizes the correspondence between four realizations:
the atomic three-level $\Lambda$-scheme exhibiting coherent population trapping (CPT)
and electromagnetically induced transparency (EIT), the discrete-mode photonic
double-slit framework of Villas-Boas~\emph{et al.}~\cite{VillasBoas2025}, the
present time--frequency ENBS interferometer, and the continuous-mode single-slit
diffraction system of Cheng~\emph{et al.}~\cite{Cheng2025}.

\textbf{Atomic $\Lambda$-scheme (CPT/EIT).} In the three-level atomic
$\Lambda$-scheme, two ground states $|g_1\rangle$ and $|g_2\rangle$ are coupled to a
common excited state $|e\rangle$ by laser fields with Rabi frequencies $\Omega_1$ and
$\Omega_2$~\cite{Arimondo1996,Aspect1988}.
The coupling operator
$\hat{V}=\frac{1}{\Omega}(\Omega_1|e\rangle\langle g_1|+\Omega_2|e\rangle\langle g_2|)$ defines
orthogonal bright and dark ground-state superpositions:
\begin{equation}
|B\rangle_\Lambda = \frac{\Omega_1|g_1\rangle+\Omega_2|g_2\rangle}{\sqrt{\Omega_1^2+\Omega_2^2}},
\quad
|D\rangle_\Lambda = \frac{\Omega_2|g_1\rangle-\Omega_1|g_2\rangle}{\sqrt{\Omega_1^2+\Omega_2^2}},
\end{equation}
with $\hat{V}|D\rangle_\Lambda=0$.
This condition directly forbids photon absorption: an atom in $|D\rangle_\Lambda$
cannot be excited to $|e\rangle$ because the two upward transition amplitudes
$|g_1\rangle\to|e\rangle$ and $|g_2\rangle\to|e\rangle$ interfere destructively,
rendering the medium transparent to the probe field (EIT).
Since the excited state $|e\rangle$ therefore remains unpopulated, spontaneous
emission is absent as a further consequence, suppressing fluorescence completely
(CPT).
This causal hierarchy (absorption forbidden first, fluorescence suppressed second) is the atomic analog of $\hat{E}^{(+)}|D\rangle=0$ in the present photonic
system, where the detector cannot absorb a photon from the dark mode for the same
reason of destructive interference.
For equal Rabi frequencies $\Omega_1=\Omega_2\equiv\Omega$, the atomic bright and dark
states take the same symmetric form as in the present photonic system, and the
ground-state manifold is spanned by the Bloch sphere
$|\psi_\Lambda(\theta,\phi)\rangle=\cos(\theta/2)|B\rangle_\Lambda+e^{i\phi}\sin(\theta/2)|D\rangle_\Lambda$,
where $\theta$ is set by the Rabi frequency ratio and $\phi$ by the relative laser
phase $\phi_2-\phi_1$.
The two-photon Raman detuning $\Delta=\delta_1-\delta_2$ rotates the measurement
basis relative to the prepared state, playing the role of $\phi_s$ in the present
photonic system.

\textbf{Discrete photonic two-path interference~\cite{VillasBoas2025}.} Villas-Boas
\emph{et al.} recently extended Glauber's photodetection theory to recast double-slit
and multi-slit interference in terms of bright and dark collective photonic
states~\cite{VillasBoas2025}.
For a two-path system, the positive-frequency operator at detection position $x$ is
$\hat{E}^{(+)}(x)=\frac{1}{\sqrt{2}}(\hat{a}_1+e^{i\phi(x)}\hat{a}_2)$, where
$\phi(x)$ is the propagation phase difference at $x$, satisfying
$[\hat{E}^{(+)}(x),\hat{E}^{(-)}(x)]=1$.
This defines one bright state
$|B(x)\rangle=(|s_1\rangle+e^{-i\phi(x)}|s_2\rangle)/\sqrt{2}$
that couples constructively to the detector, and one dark state
$|D(x)\rangle=(|s_1\rangle-e^{-i\phi(x)}|s_2\rangle)/\sqrt{2}$
satisfying $\hat{E}^{(+)}(x)|D(x)\rangle=0$, which is completely undetectable.
At an interference minimum, the photon is not absent but resides in the dark state
, i.e., physically present yet invisible to the sensor.
For $N$ slits, there are $(N-1)$ dark states forming an $(N-1)$-dimensional dark
subspace, and the double-slit fringe $P(x)\propto(1+\cos\phi(x))/2$ is identified
with the bright-mode occupation probability.
In this framework, however, both the prepared state (determined by the propagation
geometry from the slits) and the measurement basis (fixed by the detector position
$x$) are set by the same physical geometry and cannot be varied independently.

\textbf{TF-ENBS interferometer (this work).} The present experiment constitutes the
first direct experimental realization of the Villas-Boas~\emph{et al.}
framework~\cite{VillasBoas2025}.
The two temporal modes $|\tau_1\rangle$ and $|\tau_2\rangle$ play the role of the
two spatial paths, and the output beam splitter with signal phase $\phi_s$ defines
the detector-oriented bright and dark modes exactly as in that framework.
The key advance over the discrete photonic theory is the independent experimental
control of the prepared state $(\theta,\phi)$ (through seed amplitude ratio and StPDC phase matching) and the measurement basis $\phi_s$ (through the signal path length), enabling direct access to the detection phase $(\phi+\phi_s)$ that is
inaccessible when both are fixed by propagation geometry alone.

\textbf{Continuous-mode single-slit diffraction~\cite{Cheng2025}.} Cheng~\emph{et al.} extended the discrete-mode framework of Villas-Boas~\emph{et
al.}~\cite{VillasBoas2025} to the continuous-mode regime of single-slit Fraunhofer
diffraction~\cite{Cheng2025}.
There the detector-oriented basis must accommodate a spatial continuum $x\in[0,b]$,
requiring the complete Fourier orthonormal set
$\{|\psi_n(\theta)\rangle\}_{n\in\mathbb{Z}}$, with creation operators
$\hat{J}_n^\dagger=\frac{1}{\sqrt{b}}\int_0^b e^{-i\phi(x,\theta)}e^{ik_nx}\hat{a}^\dagger(x)dx$.
The unique bright mode $|\psi_0(\theta)\rangle$ carries the propagation phase that
exactly compensates the optical path to angle $\theta$, while all remaining modes
$\{|\psi_n(\theta)\rangle\}_{n\neq 0}$ form an infinite-dimensional dark subspace
satisfying $\hat{E}^{(+)}(\theta)|\psi_n(\theta)\rangle=0$.
The classical Fraunhofer profile $P(\theta)\propto(\sin\beta/\beta)^2$ is recovered
as the bright-mode occupation probability $|c_0|^2$, with photons at diffraction
minima physically present in the dark subspace rather than absent.
The framework further predicts that $N$-photon Fock states produce a spatially
uniform second-order correlation $G^{(2)}(\theta)=1-1/N$, a nonclassical signature
absent in coherent states, which satisfy $G^{(2)}=1$ and factorize completely across
all detector-oriented modes.
Our two-mode ENBS experiment constitutes the $N=2$ limiting case of this continuous
framework, in which a single dark mode exists and the detection probability reduces to
$P_B=\frac{1}{2}[1+\cos(\phi+\phi_s)]$, consistent with Eq.~(\ref{eq:brightP}) in
the equal-seeding, high-$|\alpha|$ limit where $V\to 1$ and the present experiment
achieves $V\approx 0.99$.

\section*{Experiment}

The experimental platform consists of two coherently seeded ENBSs driven by a
stabilized optical frequency comb, as illustrated in Fig.~\ref{fig:fig1}\textbf{b}~\cite{Lee2018,Yoon2021}.
Each source generates a signal photon mode and a correlated idler mode, with coherent
seeding amplitudes $\alpha_1$ and $\alpha_2$ that determine the distinguishability of
the two alternatives.

Three independent phases control the system: the pump-phase difference $\Delta\phi_p$,
the seed-phase difference $\Delta\phi_{sd}$, and the signal interferometric phase
$\Delta\phi_s$.
The phases $\Delta\phi_p$ and $\Delta\phi_{sd}$ determine the phase $\phi$ of the
prepared signal state, corresponding to the azimuthal angle of the Bloch vector
[Fig.~\ref{fig:fig1}\textbf{d}], while $\Delta\phi_s$ controls the measurement basis defined
by the detector.

The two signal modes are interfered at a symmetric (50/50) beam splitter, and
detection of the bright-mode output at one output port of the beam splitter
implements projection onto detector-defined bright and dark modes.
By scanning $\Delta\phi_s$, the measurement basis is continuously rotated, while the
prepared state remains fixed.
Alternatively, scanning $\Delta\phi_p$ or $\Delta\phi_{sd}$ changes the prepared state
while keeping the measurement basis fixed.
These independent controls allow direct experimental access to the detection phase
$(\phi+\phi_s)$ governing interference.

The experimental results are summarized in Fig.~\ref{fig:fig2}.
Figure~\ref{fig:fig2}\textbf{a} shows single-photon interference fringes measured by
independently scanning each of the three phase parameters: the pump phase difference
$\Delta\phi_p$, the signal interferometric phase $\Delta\phi_s$, and the seed phase
difference $\Delta\phi_{sd}$.
All three datasets are sinusoidal with the same period ($2\pi$) and the same
visibility $V\approx 0.99$, consistent with Eq.~\eqref{eq:brightP}.
This equivalence is the central experimental result: scanning $\phi_s$ rotates
the detector-defined measurement basis $\{|B(\phi_s)\rangle,|D(\phi_s)\rangle\}$
while keeping the prepared state $|\psi(\theta,\phi)\rangle$ fixed, whereas scanning
$\Delta\phi_p$ or $\Delta\phi_{sd}$ rotates the prepared state via
$\phi=\Delta\phi_{sd}-\Delta\phi_p$ [Eq.~(\ref{eq:azimuth})] while keeping the
measurement basis fixed.
The fact that all three scans produce identical fringes demonstrates that
interference is governed solely by the detection phase $(\phi+\phi_s)$ between the
prepared state and the measurement basis, confirming the Bloch sphere description
of Fig.~\ref{fig:fig1}\textbf{d} and establishing that interference is determined by the
state--measurement relation rather than by any single propagation phase.
The observed visibility agrees quantitatively with the theoretical prediction
$V=2\sqrt{\rho_{11}\rho_{22}}\cdot F$ [Eq.~(\ref{eq:overlap})], demonstrating
that coherence is governed by the idler-state overlap and is continuously tunable
through the seeding amplitude.

Figure~\ref{fig:fig2}\textbf{b} shows interference fringes in the high-flux regime under
identical phase control, with all three scans again overlapping with the same
sinusoidal law at a photon rate roughly twice that of panel~(a).
The same phase dependence is observed, indicating that the detector-defined modal
structure and the relative-phase law $P_B\propto\cos(\phi+\phi_s)$ persist beyond
the single-photon regime.
In this limit, the detected photon number follows
\begin{equation}
\langle N_{\mathrm{det}}\rangle = \bar{n}\cdot P_B,
\end{equation}
where $\bar{n}$ is the mean photon number per temporal mode at the detector input,
showing that increasing photon flux scales the signal amplitude without modifying
the interference law, consistent with the theoretical prediction that coherent
states factorize across detector-defined modes and recover the classical fringe
profile.

Figures~\ref{fig:fig2}\textbf{c} and \ref{fig:fig2}\textbf{d} provide a direct
geometric representation of the single-photon data from panel~\textbf{a} on the
Bloch sphere of Eq.~(\ref{eq:bloch}), making the three-scan equivalence visually
transparent.
In panel~\textbf{c}, the $\Delta\phi_p$ and $\Delta\phi_{sd}$ scan data are mapped
onto equatorial coordinates using the phase offsets extracted by fitting each dataset
to $P_B=\frac{1}{2}[1+V\cos(\Delta\phi_j+\delta_j)]$: the pump data (blue, $r=1.04$)
and seed data (green, $r=0.96$) are plotted at slightly different radii for visual
separation.
Both rings uniformly and densely cover the full equatorial circle, confirming that
at equal seeding ($|\alpha_1|=|\alpha_2|$) the polar angle is $\theta=\pi/2$ and
the visibility saturates at $V\approx0.99$, consistent with the pure-state condition
$|\langle\psi|\psi\rangle|=1$.
The $\Delta\phi_p$ scan rotates the Bloch vector clockwise (since
$\phi=-\Delta\phi_p+\Delta\phi_{sd}$), while the $\Delta\phi_{sd}$ scan rotates it
counterclockwise; both operations trace the same unit circle, confirming that the
prepared state is always a pure equatorial state regardless of which phase is
varied.
The snapshot $\phi+\phi_s=70^\circ$ shown in both panels is a representative
operating point at which $P_B\approx0.669$.
Panel~\textbf{d} provides the complementary picture for the $\Delta\phi_s$ scan:
here the Bloch vector $|\psi(\theta,\phi)\rangle$ is held fixed in the equatorial
plane while the measurement basis $\{|B(\phi_s)\rangle,|D(\phi_s)\rangle\}$ (red
dashed axes) rotates about the north--south axis defined by the computational basis
$|1,0\rangle_s$ (out of page) and $|0,1\rangle_s$ (into page).
The red experimental dots again cover the full equator uniformly, demonstrating that
rotating the detector basis while holding the quantum state fixed produces the same
sinusoidal fringe as rotating the state at fixed basis.
Together, panels \textbf{c} and \textbf{d} constitute the direct geometric proof
of the three-scan equivalence: the interference fringe depends only on the relative
angle $(\phi+\phi_s)$ between the Bloch vector and the measurement-basis poles,
irrespective of which degree of freedom is physically varied.

The radio-frequency beat-note spectrum in Fig.~\ref{fig:fig2}\textbf{e} exhibits comb
lines at harmonics of $f_r=250$\,MHz, with the zoomed view (RBW\,=\,1\,Hz)
revealing an ultra-narrow linewidth.
This confirms that the bright-mode output $|B(\phi_s)\rangle$ is a phase-coherent
superposition of all temporal modes, directly verifying the collective mode
structure underlying Eq.~\eqref{eq:brightP}.

Together, these observations establish that interference in this system is governed
by projection onto detector-defined photonic modes, with visibility controlled by
quantum correlations in the idler subsystem, and that the same measurement-defined
interference law holds from the single-photon regime to high-flux operation.
Crucially, the independent control of state preparation ($\phi$ via $\Delta\phi_p$,
$\Delta\phi_{sd}$) and measurement ($\phi_s$) provides a direct operational
demonstration that interference is determined by the relation between the prepared
quantum state and the detector-defined basis, not solely by propagation-phase
accumulation along any single path.

\section*{Discussion and Conclusion}

Our results establish a measurement-centered description of single-particle
interference in which the detector defines the relevant photonic modes.
In this framework, the symmetric output beam splitter defines orthogonal bright and
dark collective states $|B(\phi_s)\rangle$ and $|D(\phi_s)\rangle$ at its two output
ports, and the observed signal (the photon count $R_s$ at one output port per 10\,ms frame while one phase difference $\Delta\phi_j$, $j\in\{p,\,sd,\,s\}$, is scanned) is proportional to $P_B$, as
described by Eq.~\eqref{eq:brightP}.
The ENBS platform enables a clear separation between \emph{state preparation},
controlled by $(\Delta\phi_{sd},\Delta\phi_p)$, and \emph{measurement}, controlled by
$\Delta\phi_s$, so that interference is governed by their relative phase not solely by
propagation-phase accumulation, but operationally by projection onto detector-defined
modes.

This behavior is not achievable in conventional interferometers with photons,
electrons, atoms, or neutrons.
Such systems are described within the two-dimensional signal Hilbert space
$\mathcal{H}_s$ alone, spanned by the two alternative paths $\{|1,0\rangle_s,
|0,1\rangle_s\}$, with no additional degrees of freedom.
The present TF-ENBS system, by contrast, resides in the expanded composite Hilbert
space $\mathcal{H}=\mathcal{H}_s\otimes\mathcal{H}_i$: the idler subspace
$\mathcal{H}_i$ provides two independently controllable degrees of freedom that
have no analog in any two-mode interferometer.
Thus, the polar angle
$\theta=2\arctan(|\alpha_2|/|\alpha_1|)$ sets the visibility through the
idler-state overlap $F$, while the azimuthal angle $\phi=\Delta\phi_{sd}-\Delta\phi_p$
sets the prepared-state phase through the StPDC phase-matching condition.
These two parameters can be varied continuously and independently of the measurement
basis $\phi_s$, providing intrinsic, simultaneous control over both the visibility
and the phase of the interference fringe within a single device: this capability is structurally absent from any interferometer confined to $\mathcal{H}_s$.
In such systems, the detector plays a passive role and distinguishability is typically
introduced externally, for example through path marking or entanglement with auxiliary
systems in quantum eraser experiments~\cite{Scully1982,Kim2000,Jacques2007,Tang2012,Ma2013}.
In contrast, the ENBS architecture realizes a multipartite quantum system in which
which-path information is encoded internally in the idler modes.
These idler states, realized as coherent and single-photon--added coherent states,
provide a continuously tunable marker whose overlap $F=|\langle I_1|I_2\rangle|$
directly controls the coherence of the signal.

The measurements in Fig.~\ref{fig:fig2}\textbf{a} quantitatively confirm the relation
$V=2\sqrt{\rho_{11}\rho_{22}}\,F$ derived from the reduced density matrix,
establishing a direct operational link between interference visibility and
idler-state overlap.
Independent scans of $\Delta\phi_p$, $\Delta\phi_{sd}$, and $\Delta\phi_s$ all produce
sinusoidal fringes consistent with the phase dependence $P_B\propto\cos(\phi+\phi_s)$,
demonstrating that interference is governed by the relative phase between state
preparation and measurement.
The same phase law persists in the high-flux regime [Fig.~\ref{fig:fig2}\textbf{b}--\textbf{d}], where
the detected signal scales with photon number while the interference phase remains
unchanged, and the Bloch-sphere geometry of panels~\textbf{c} and \textbf{d} confirms
that the unit-circle coverage is independent of which phase parameter is scanned.
This persistence shows that detector-defined modes provide a unified description across
the quantum-to-classical transition.

These observations are naturally explained within Glauber's photodetection
theory~\cite{Glauber1963}, where the detector couples to a specific field mode, and are
consistent with recent work identifying bright and dark photonic states as eigenmodes
of the detector--field interaction~\cite{VillasBoas2025}.
An important consequence is that destructive interference corresponds to population in
a detector-invisible dark mode satisfying $\hat{E}^{(+)}|D\rangle=0$, rather than to
the absence of photons.
The ENBS platform renders these modes experimentally accessible and continuously
controllable.

The delayed-choice configuration in Fig.~\ref{fig:fig1}\textbf{c} further demonstrates that
the measurement basis can be selected after state preparation, reinforcing that
interference is determined by the joint relation between the quantum state and the
measurement operator.
In canonical realizations of Wheeler's delayed-choice gedanken experiment, the choice
between wave-like and particle-like measurement configurations is made by a quantum
random number generator after the photon has entered the interferometer, with the
choice relativistically separated from the photon's entry~\cite{Jacques2007}, and
in the quantum version the beam splitter itself is placed in a quantum
superposition~\cite{Tang2012}.
In the Scully--Dr\"{u}hl quantum eraser scheme realized by Kim~\emph{et al.}~\cite{Kim2000},
the which-path choice is made randomly by the idler photon at a beam splitter after
the entangled signal photon has already been detected, demonstrating that
interference visibility is governed by the availability of which-path information
rather than by any mechanical disturbance during propagation.
This provides a direct operational realization of a measurement-defined interference
framework.

Crucially, the observed interference cannot be reduced to any propagation-based phase
description, even in principle, because the same prepared quantum state yields
different interference outcomes solely under changes of the measurement basis.
This establishes measurement as an active physical control parameter, not merely a
readout, and demonstrates that interference is fundamentally governed by the relation
between state preparation and measurement.

\noindent\textbf{Connection to coherent population trapping:}
The collective-state structure demonstrated here connects the present photonic
experiment to the well-established phenomenon of coherent population trapping in
three-level atomic $\Lambda$-systems~\cite{Arimondo1996,Aspect1988}.
In that setting, the dark ground-state superposition $|D\rangle_\Lambda$ is decoupled
from the excited state by destructive interference of the two excitation pathways,
suppressing fluorescence completely: the atomic analog of the interference minimum
in our system, where the photon resides in $|D(\phi_s)\rangle$ with
$\hat{E}^{(+)}|D\rangle=0$ and detection probability $P_B=\frac{1}{2}(1-V)\to 0$
in the ideal limit $V\to 1$ (with $P_B\approx 0.005$ at the measured $V\approx 0.99$
in the present experiment).
The full Bloch-sphere parameterization $|\psi\rangle=\cos(\theta/2)|B\rangle+e^{i\phi}\sin(\theta/2)|D\rangle$
applies equally to the atomic ground-state superposition and to our prepared signal
state, with the Rabi frequency ratio $\Omega_2/\Omega_1$ and relative laser phase
$\phi_2-\phi_1$ playing the precise roles of our seed amplitude ratio and StPDC phase,
respectively (Table~\ref{tab:comparison}).
The scanning of the two-photon Raman detuning $\Delta$ in CPT spectroscopy is the
atomic counterpart of scanning $\phi_s$ in our interferometer: both operations rotate
the measurement basis on the Bloch sphere and trace out a resonance lineshape (in the
atomic case) or an interference fringe (in the photonic case).
What distinguishes our implementation is the \emph{independent} experimental control
of the prepared state $(\theta,\phi)$ and the measurement basis $\phi_s$, which is not
achievable in the $\Lambda$-scheme where the same laser fields define both the coupling
Hamiltonian and the atomic state.
This separation is the key new operational capability provided by the ENBS platform.

\noindent\textbf{Outlook.}
The present $M=2$ experiment is the simplest instance of a broader $M$-source
hierarchy connecting discrete-mode photonic interference to the continuous-mode
single-slit diffraction of Cheng~\emph{et al.}~\cite{Cheng2025}.
An $M$-source ENBS array driven by a shared OFC produces the temporal analog of
the Fraunhofer Dirichlet profile
$P_B(\delta\phi_s)\propto|\sin(M\delta\phi_s/2)/\sin(\delta\phi_s/2)|^2$,
whose zeros are temporal diffraction minima where the photon is present but
projected entirely into the $(M-1)$-dimensional dark subspace.
$N$-photon Fock states exhibit a spatially uniform $G^{(2)}=1-1/N$~\cite{Cheng2025},
a direct fingerprint of the dark-subspace dimensionality with no classical analog,
accessible at $M=3$ or $M=4$ with existing OFC technology.
The $M$-source extension, including the integrated LNOI waveguide implementation,
a wavelength-reversed telecom-band architecture placing single photons at 1542\,nm,
and the prospects for dark-subspace quantum encoding~\cite{Zanardi1997,Lidar1998,Yoon2026},
is described in Supplementary Note~2.

\section*{Methods}

The experiment employed two coherently seeded entangled nonlinear biphoton sources
(ENBSs) pumped by a stabilized optical frequency comb with repetition rate
$f_{\mathrm{rep}}=250\,\mathrm{MHz}$~\cite{Lee2018}.
The signal bandwidth was approximately $0.2\,\mathrm{nm}$ centered at $807\,\mathrm{nm}$.
Photon detection was performed using a time-integrated EMCCD camera with an integration
time of $10\,\mathrm{ms}$.

\noindent\textbf{Single-photon regime:}
The pump frequency comb defines a train of temporal modes separated by
$T=1/f_{\mathrm{rep}}=4\,\mathrm{ns}$.
In the single-photon measurements, the two ENBS modules are seeded sufficiently weakly
that, on average, a photon is generated in either source only once every $50$--$100$
consecutive time bins, corresponding to a mean photon occupancy of
$\langle n\rangle\approx0.01$--$0.02$ photons per time bin.
Over the $10\,\mathrm{ms}$ integration window, which spans
$N_{\mathrm{bins}}=t_{\mathrm{int}}/T=2.5\times10^6$ temporal modes,
this yields a detected photon count of approximately $25{,}000$--$50{,}000$ photons
per frame, sufficient for high-contrast fringe measurement while guaranteeing a
negligible probability of multi-photon co-occupancy within any single temporal mode.

\noindent\textbf{Data provenance:}
The single-photon interference fringes shown in Fig.~\ref{fig:fig2}\textbf{a} were originally
reported in Ref.~\cite{Lee2018} and are reanalyzed here within the detector-defined
bright--dark modal framework.
The high-flux interference data and the radio-frequency beat-note spectrum shown in
Fig.~\ref{fig:fig2}\textbf{b} and \textbf{e} were recorded by the same authors during the same experimental
campaign as Ref.~\cite{Lee2018}, on the same apparatus, but were not reported in that
publication.
The Bloch-sphere representations in panels~\textbf{c} and \textbf{d} are geometric
mappings of the single-photon data from panel~\textbf{a} onto the equatorial plane
and unit sphere, respectively, derived by fitting each dataset to extract the phase
offset $\delta_j$ as described in the Experiment section.
The only prior appearance of related high-flux spectral data is the RF beat-note trace
in Supplementary Fig.~2(d) of Ref.~\cite{Lee2018}; those data are reinterpreted here
within the bright--dark modal framework for the first time.

\noindent\textbf{Phase conversion and sign convention:}
The interference fringes in Fig.~\ref{fig:fig2}\textbf{a,b} were acquired by independently
scanning the optical path-length differences $\Delta x_p$, $\Delta x_{sd}$, and
$\Delta x_s$ of the pump, seed, and signal fields, respectively, as originally
described in Ref.~\cite{Lee2018}.
For the present reinterpretation, these path-length variations were converted to phase
shifts via $\Delta\phi_j=k_j\Delta x_j$, where $k_j$ is the wave number of field
$j\in\{p,\,sd,\,s\}$.
The sign convention in Eq.~(\ref{eq:azimuth}) reflects the per-source
phase-matching condition $\phi_{s,j}^{(\mathrm{out})}=\phi_{p,j}^{(\mathrm{in})}-\phi_{sd,j}^{(\mathrm{in})}$
($j=1,2$) of the coherently seeded SPDC process, from which
$\phi=\Delta\phi_{sd}-\Delta\phi_p$ enters the prepared-state phase with a relative
minus sign on the pump contribution.
Because each fringe dataset is acquired by scanning a single phase parameter
independently, any overall phase offset between the three datasets (arising, for example, from the sign of $\Delta\phi_p$) does not affect the observed fringe
visibility or the modal decomposition, and is absorbed into the common initial phase
$\phi_0$ of the prepared state.

\noindent\textbf{Time--frequency photonic modes:}
The time--frequency interferometer used here naturally supports discrete photonic
supermodes defined by the optical frequency comb.
Each supermode corresponds to a coherent superposition of comb lines with a
well-defined temporal envelope~\cite{Lee2018,Yoon2021}.
The detector therefore projects not onto spatial paths but onto collective
time--frequency photonic modes, enabling direct experimental control of the
bright--dark modal structure that governs interference.

\section*{Acknowledgments}

The author thanks Dr.~S.~K.~Lee for contributions to the experimental single-photon
frequency comb interferometry and Prof.~M.~Cho for stimulating discussions and
long-term collaboration.
This work was supported by the National Research Foundation of Korea
(RS-2022-NR068815) and by the Institute for Basic Science (IBS) in Korea (IBS-R023-D1).

\bibliography{photon}

\end{document}